\documentclass[aps,prl,twocolumn,showpacs,superscriptaddress,floatfix]{revtex4-1}

\usepackage{braket,mathtools,graphicx,siunitx}
\usepackage[colorlinks=true, allcolors=blue]{hyperref}

\begin{document}
\title{Generating maximal entanglement between spectrally distinct solid-state emitters}

\author{David L. Hurst$^\dagger$}
\affiliation{Physics \& Astronomy, University of Sheffield, Hounsfield Road, Sheffield, S3 7RH, United Kingdom}

\author{Kristoffer B. Joanesarson$^\dagger$}
\affiliation{Physics \& Astronomy, University of Sheffield, Hounsfield Road, Sheffield, S3 7RH, United Kingdom}
\affiliation{Department of Photonics Engineering, Technical University of Denmark, \O rsteds Plads, Lyngby, Denmark}

\author{Jake Iles-Smith}
\affiliation{Physics \& Astronomy, University of Sheffield, Hounsfield Road, Sheffield, S3 7RH, United Kingdom}

\author{Jesper M\o rk}
\affiliation{Department of Photonics Engineering, Technical University of Denmark, \O rsteds Plads, Lyngby, Denmark}

\author{Pieter Kok}\email{p.kok@sheffield.ac.uk}
\affiliation{Physics \& Astronomy, University of Sheffield, Hounsfield Road, Sheffield, S3 7RH, United Kingdom}
\date{\today}

\begin{abstract}
 \noindent  
 We show how to create maximal entanglement between spectrally distinct solid-state emitters embedded in a waveguide interferometer. 
 By revealing the rich underlying structure of multi-photon scattering in 
 emitters,
 we show that a two-photon input state can generate deterministic maximal entanglement even for emitters with significantly different transition energies and line-widths. The optimal frequency of the input is determined by two competing processes: which-path erasure and interaction strength.  We find that smaller spectral overlap can be overcome with higher photon numbers, and quasi-monochromatic photons are optimal for entanglement generation. 
 Our work provides a new methodology for solid-state entanglement generation, where the requirement for perfectly matched emitters can be relaxed in favour of optical state optimisation.

\end{abstract}

\maketitle


\noindent
Quantum technologies promise dramatic improvements in computing and communication by utilizing quantum entanglement between qubits \cite{nielsen2002quantum}. Although many promising quantum technology architectures have emerged over the last two decades, none are free from the practical challenges presented by high-fidelity quantum control and scalability. For example, superconducting circuit implementations enjoy excellent coherence properties but operate slowly~\cite{Wendin17}, while trapped ion qubits can be prepared with almost unit fidelity but are difficult to scale~\cite{Kielpinski02}. Solid state architectures, such as optically coupled spin systems, compete on speed and scalability. They include semiconductor quantum dots and nitrogen-vacancy centres. Large optical non-linearities in solid-state systems are now very common \cite{Hallett18,Javadi15,Sipahigil16}, and solid-state emitters are readily integrated into complex photonic structures, further enhancing the light-matter interaction~\cite{Lodahl15}. However, there are many challenges still to overcome. For example, charge noise and phonon scattering have limited the size of the optical non-linearities observed thus far \cite{Hallett18}.

Another major drawback to solid-state emitters is that the central energies and lifetimes of their transitions are highly dependent on the fabrication process, and vary significantly both across and within samples~\cite{Michler09}. Known methods for entangling solid-state qubits require emitters with identical energies to facilitate path-erasure techniques \cite{Barrett05,Metz08}. This adds a practically insurmountable overhead to the process of matching multiple solid-state qubits for creating large entangled states~\cite{Raussendorf01}. Stark shifting and strain tuning the emitter transitions has been employed to tune solid-state emitters onto resonance~\cite{Stockill17,Bernien13,Hensen15}, but this requires a substantial technical overhead and arbitrary emitters in a sample cannot in general be tuned onto resonance. 
      Here, we propose a process for generating entanglement that is robust against spectral variations in the emitters' transition energies and line-widths. We show that photons, linear optics, and photon counting suffice to create deterministic entanglement between imperfectly matched emitters, revealing a rich underlying structure of multi-photon scattering off two non-identical emitters. While many challenges remain, this work removes a major obstacle to a scalable solid state quantum technology architecture. 

\begin{figure}[t]
 \begin{center}
 \includegraphics[scale=1]{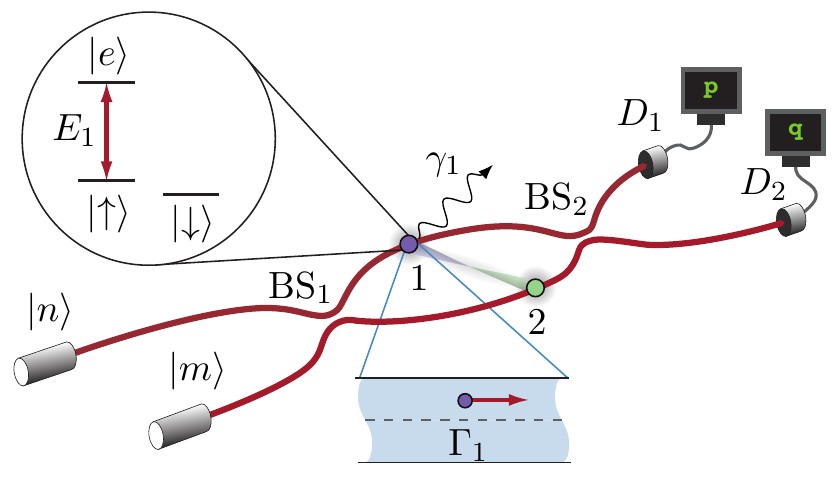}
 \caption{Waveguide Mach-Zehnder interferometer with emitters embedded at positions $1$ and $2$, and with $L$-type level structures shown in the inset. The excited state $\ket{e}$ is coupled to a spin qubit state (e.g., $\ket{\uparrow}$) with transition energy $E_\alpha$ ($\alpha=1,2$), circular polarisation, and line-width $\Gamma_\alpha$. The emitters are placed off-axis in the waveguide at c-points, such that circularly polarised light scatters only in the forward direction. The loss rate from the guided mode is $\gamma_\alpha$. Fock states $\ket{n,m}$ are injected into the interferometer, and detectors $D_1$ and $D_2$ record a photon number detector signature $(p,q)$.} 
 \label{fig:mzi}
 \end{center}
\end{figure}


Our setup is shown in Fig.~\ref{fig:mzi}. Two solid-state emitters each have an $L$-type level structure, with two stable low-lying spin states ($\ket{\uparrow}$, $\ket{\downarrow}$), and a dipole transition that couples a spin state to an excited state $\ket{e}$. The transition energy for emitter $\alpha=1,2$ is $E_\alpha$, and the polarisation is circular due to selection rules. The emitters are initially prepared in the product state $(\ket{\uparrow} + \ket{\downarrow})(\ket{\uparrow} + \ket{\downarrow})/2$, and embedded in a waveguide Mach-Zehnder interferometer at c-points, where perfect correlation between propagation direction and circular polarisation occurs \cite{Lodahl17}. Consequently, the emitters scatter circularly polarised light only in the forward direction, as was demonstrated recently using semiconductor quantum dots under an applied magnetic field~\cite{Sollner15,Coles16,Lang17,Hurst18a}. For a lossless waveguide, the emitter will impart a $\pi$-phase shift to each photon that is on resonance with the transition~\cite{Fan10,Roulet16,Nysteen17,Hurst18b}. The input to the interferometer is a two-mode Fock state $\ket{n,m}$, and the detectors $D_1$ and $D_2$ produce classical signatures $(p,q)$ indicating the presence of $p$ and $q$ photons, respectively. 
 
Assuming the emitters are identical, a single monochromatic resonant photon injected into either one of the input arms of the interferometer will scatter from one of the emitters, and after the final beam splitter the light-matter degrees of freedom are in the multi-partite entangled state $\left(\ket{\Phi^-}\otimes\ket{1,0} - \ket{\Psi^-}\otimes\ket{0,1}\right)/\sqrt{2}$, where $\ket{0,1}$ and $\ket{1,0}$ is the two-mode single photon state at the interferometer output. A photon detector signature $(1,0)$ or $(0,1)$ heralds the maximally entangled spin state $\ket{\Phi^-} = \left(\ket{\uparrow,\uparrow}-\ket{\downarrow,\downarrow}\right)/\sqrt{2}$ or $\ket{\Psi^-} = \left(\ket{\uparrow,\downarrow}-\ket{\downarrow,\uparrow}\right)/\sqrt{2}$. This method can be made robust to photon loss provided that the emitter detuning is only a fraction of the emitter line-width. Mahmoodian \emph{et al}.~showed how this can form a building block for distributed quantum computing~\cite{Mahmoodian16}.

In practice, both the line-widths and transition energies vary significantly between solid-state emitters, and it was generally assumed that this prohibits the creation of perfect entanglement using linear optics and photodetection. In this case, the input photon can no longer be resonant with both emitters simultaneously. With $\hbar\omega$ the single-photon energy, $\Gamma_\alpha$ the unidirectional emission rate of emitter $\alpha=1,2$ into the waveguide, and $\gamma_\alpha$ the corresponding coupling to non-guided modes, the scattering process is described by the transmission coefficient \cite{Rephaeli13}
\begin{align}
 t_\alpha (\omega) = \frac{\hbar\omega-E_\alpha - i\hbar(\Gamma_\alpha- \gamma_\alpha)/2}{\hbar\omega-E_\alpha + i\hbar(\Gamma_\alpha + \gamma_\alpha)/2}\, .
\label{eq:trans}\end{align}
We characterise the emitter loss by $\beta_\alpha\equiv\Gamma_\alpha/(\Gamma_\alpha+\gamma_\alpha)$. For non-zero emitter detuning $\delta\equiv E_2-E_1$, $t_\alpha (\omega)$ ceases to be a $\pi$ phase shift, and for $\beta_\alpha<1$, $t_\alpha (\omega)$ is no longer a pure phase shift. The setup then does not create maximally entangled states deterministically anymore. Nevertheless, we will now demonstrate how tailoring the optical input state $\ket{n,m}$ into the Mach-Zehnder interferometer leads to deterministic maximal entanglement between two spectrally distinct emitters.

In general, a detector signature $(p,q)$ indicates that the two emitters are in a mixed entangled state. We use the concurrence $\mathcal{C}(\rho)$ for a two-qubit state $\rho$ to quantify this entanglement~\cite{Wootters98}. Each signature $(p,q)$ occurs with probability $\text{Pr}(p,q)$ and results in an emitter state $\rho_{(p,q)}$, leading to a concurrence $\mathcal{C}({\rho_{(p,q)}})$. We define the \emph{average} concurrence as 
\begin{align} 
 \mathcal{C}_\text{avg} \equiv \sum_{(p,q)} \text{Pr}(p,q)\, \mathcal{C}\left(\rho_{(p,q)}\right)\, .
\end{align} 
This is an appropriate figure of merit, since it provides a lower bound for the amount of entanglement expected from a given experiment without post-selection. The entanglement in the two-qubit state can be increased by discarding measurement outcomes corresponding to below-average concurrences. This comes at the expense of the rate of entanglement generation.

\begin{figure}[t]
\begin{center}
\includegraphics[scale=1]{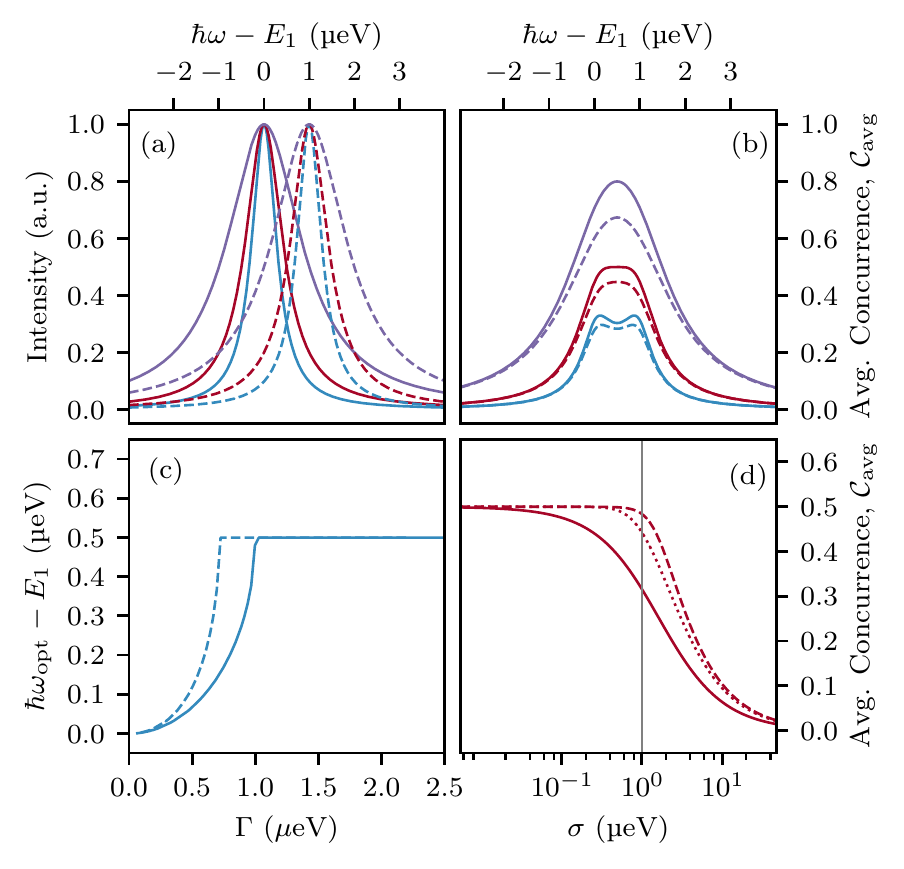}
\caption{Single-photon (e.g., $\ket{1,0}$) entanglement generation for a pair of detuned L-type emitters with equal line-width $\Gamma$ and energies $E_1$ and ${E_2=E_1+\delta}$, where $\delta=1.0$~\si{\micro}eV. (a) Lorentzian spectra for emitters with energies $E_1$ (solid), $E_2$ (dashed), and emitter line-widths $0.66$~\si{\micro}eV (blue), $1.0$~\si{\micro}eV (red), and $2.0$~\si{\micro}eV (purple). (b) Average concurrence versus monochromatic single-photon energy without loss ($\beta = \beta_1 = \beta_2 = 1$, solid) and with loss ($\beta = 0.9$, dashed). Line colours as in (a). (c) Location of optimum single-photon energy $\hbar\omega_{\rm opt}$ for maximum $\mathcal{C}_\text{avg}$ as a function of emitter line-width $\Gamma$ for a monochromatic input without loss (solid) and with $\beta = 0.9$ (dashed). (d) $\mathcal{C}_\text{avg}$ for Lorentzian (solid), Gaussian (dotted), and square (dashed) single-photon envelopes as a function of FWHM pulse-width $\sigma$. Here, $\hbar\omega=(E_1+E_2)/2$ and $\Gamma=1.0$~\si{\micro}eV. The vertical line indicates the line-width of the emitters.} 
\label{fig:fig2}
\end{center}
\end{figure}

The amount of entanglement that can be generated between the two spectrally distinct emitters with a single probe photon is shown in Fig.~\ref{fig:fig2}. 
The single photon protocol is analysed using linear optics transformations~\cite{Mahmoodian16}, while a multiphoton input requires taking into account the non-linear nature of the interaction~\cite{Nysteen17} (see SI for details). 
As expected, for spectrally distinct emitters the average concurrence does not reach its maximal value [Fig.~\ref{fig:fig2}(b)]. The amount of entanglement is determined by two competing processes. On the one hand, which-path information for the probe photon must be erased, while at the same time the phase shift induced by the photon scattering event must be maximised. Tuning closer to either emitter increases the relative phase shift but also imparts a degree of path information onto the probe, as the light-matter interaction is now stronger for one of the emitters. For emitters with finite detuning and line-width it is not obvious which photon energy maximises the average concurrence. Three emitter line-widths are shown in Fig.~\ref{fig:fig2}(a), and Fig.~\ref{fig:fig2}(b) shows the corresponding $\mathcal{C}_\text{avg}$. The linewidths shown correspond to emitters with $1$, $0.66$ and $0.33$ ns lifetimes, typical of semiconductor QDs benefiting from modest Purcell enhancements~\cite{Hughes:04}. Increasing the line-width of the emitters leads to a larger spectral overlap, thereby erasing some of the which-path information and increasing $\mathcal{C}_\text{avg}$. Fig.~\ref{fig:fig2}(c) shows the optimal frequency of the input photon that maximises $\mathcal{C}_\text{avg}$. For narrow line-widths it is preferable to tune the photon energy away from the mean emitter energy ($\hbar\omega-E_1 = 0.5$~\si{\micro}eV for $\delta = 1.0$~\si{\micro}eV), and towards resonance with one of the emitters. Though this reduces the concurrence in the state heralded by a click at detector $D_2$, it does increase the probability of a successful scattering event. 

One may expect that a photon with a wide frequency bandwidth that overlaps with both emitters will improve the entanglement generation. Fig.~\ref{fig:fig2}(d) shows the average concurrence for a single probe photon with Lorentzian, Gaussian and square spectral profiles, centred at $\hbar\omega=E_1+\delta/2$, as a function of the photon bandwidth. We find that increasing the bandwidth of the input photon only degrades the average concurrence, and a narrow-band probe is always preferable. We attribute this to the reduced temporal extent of the photon at larger bandwidths, which increases the probability of exciting the emitter, and thus the fraction of light emitted incoherently through spontaneous emission. This reduction is particularly noticeable for a Lorentzian wave-packet, where a close spectral match with the emitter increases the excitation probability. We conclude that for given emitter detuning and line-widths, the maximum $\mathcal{C}_\text{avg}$ of the single-photon case is limited by the competing requirements of maximising the induced phase shift and path erasure.

\begin{figure}[t]
 \begin{center}
 \includegraphics[scale=1]{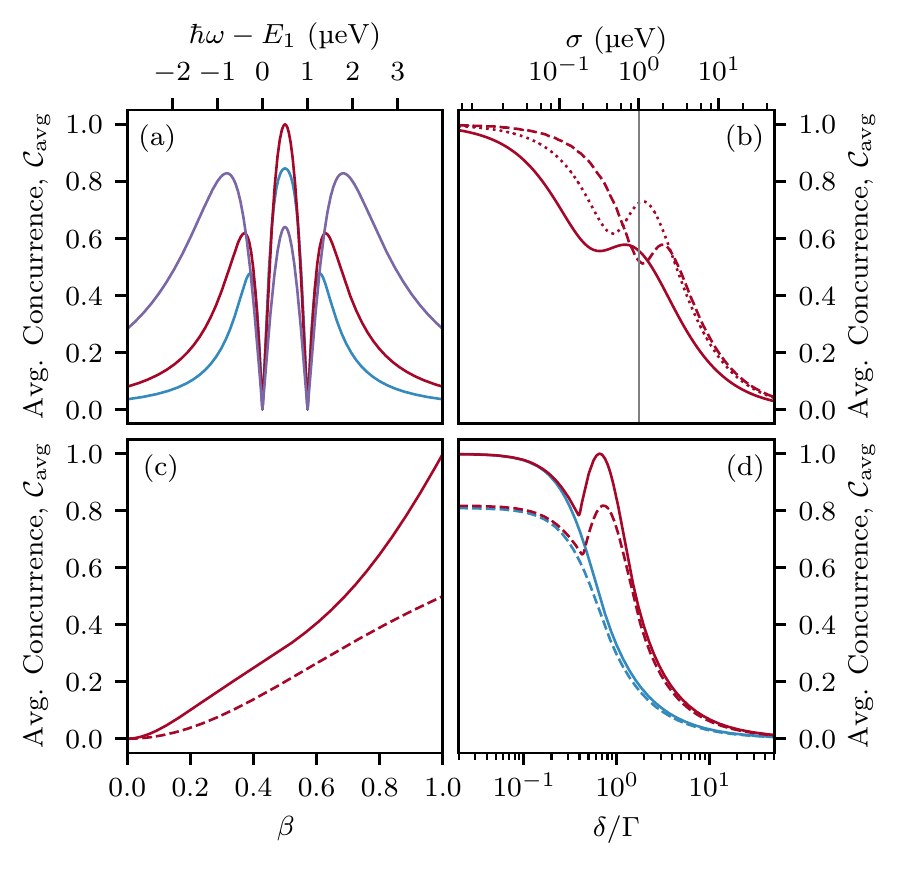}
 \caption{Two-photon (i.e., $\ket{1,1}$) entanglement generation for a pair of detuned L-type emitters with equal line-width $\Gamma$ and energies $E_1$ and ${E_2=E_1+\delta}$, where $\delta=1.0$~\si{\micro}eV. (a) Average concurrence versus monochromatic two-photon energy. The emitters have equal line-width of $0.66$~\si{\micro}eV (blue), $1.0$~\si{\micro}eV (red), and $2.0$~\si{\micro}eV (purple). (b) $\mathcal{C}_\text{avg}$ for Lorentzian (solid), Gaussian (dotted), and square (dashed) single-photon envelopes as a function of FWHM pulse-width $\sigma$. Here, $\hbar\omega=(E_1+E_2)/2$ and $\Gamma=1.0$~\si{\micro}eV. The vertical line indicates the line-width of the emitters. (c) Average concurrence as a function of $\beta=\beta_1=\beta_2$ for a monochromatic two-photon pulse (solid) and monochromatic single-photon pulse (dashed); both emitters have line-widths of $1.0$~\si{\micro}eV. (d) Average concurrence as a function of normalised emitter detuning $\delta/\Gamma$ for two monochromatic photons (red) and a single monochromatic photon (blue). Solid lines represent $\beta=1$ and dashed lines represent $\beta=0.9$.The optimal photon frequencies for various coupling and detuning ratios are discussed in the Supplementary Information.} 
 \label{fig:fig3}
 \end{center}
\end{figure}

Next, we consider whether two photons can increase the average concurrence. Consider an input state of two identical monochromatic photons $\ket{n,m}=\ket{1,1}$ entering the interferometer. They will evolve into a two-photon {\sc noon} state $(\ket{2,0}-\ket{0,2})/\sqrt{2}$ via Hong-Ou-Mandel interference on the first beam splitter~\cite{Hong87,PhysRevA.91.063823} and interact with the emitters. Entanglement is then heralded by three detector signatures: two photons in $D_1$, two photons in $D_2$, or a coincidence count. Using two probe photons leads to a rich structure in the average concurrence, and it is now possible to reach deterministic maximal entanglement for spectrally distinct emitters with finite line-width. The reason for the two-photon advantage can be determined via inspection of Fig.~\ref{fig:fig3}(a), where $\mathcal{C}_\text{avg}$ is shown as a function of the detuning between the photon energy and the transition energy of the first emitter. In the current example where $\delta = 1.0$~\si{\micro}eV, maximum entanglement fidelity occurs for emitters with line-widths of 1.0~\si{\micro}eV and input photons with energy $\hbar\omega = E_1 + \delta/2$. Comparing this value to Fig.~\ref{fig:fig2}(a), this input energy corresponds to the point where the emitter spectra are at half of their maximum intensity. 
For quasi-monchromatic input states, the imparted phase shift is additive in photon number, i.e., for the two-photon case, each photon imparts a $\pi/2$ phase shift to the emitter and therefore achieves the required $\pi$ phase shift. We consider more general emitter detuning and line-width examples in Fig.~\ref{fig:fig4} and in the Supplementary Information.

We study again the effect of broadband probe photons on the entanglement generation process. 
In this case the phase shift is no-longer additive in photon number, since the interaction becomes non-linear.
When two photons are present, the emitter may be excited, which opens a pathway for stimulated emission, such that the coherence of the wave-packet is maintained. Competition between spontaneous and stimulated emission processes leads to the non-monotonic behaviour in $\mathcal{C}_\text{avg}$ presented in Fig.~\ref{fig:fig3}(b). The two-photon interaction strength increases when the optical pulse width is broadened to the scale of the emitter line-width~\cite{Nysteen17}. 
Although an increased bandwidth generally reduces $\mathcal{C}_\text{avg}$, a local maximum appears close to the emitter line-width. We attribute this to the stimulated emission process, which is maximised for photons that are closely matched to the emitters' spectral profile. Fig.~\ref{fig:fig3}(c) shows the dependence of $\mathcal{C}_\text{avg}$ on the loss rate $\beta$. In the regime where it is preferable to tune the input photon to the average emitter energy, the effect of reducing the beta factor of either emitter causes the average concurrence to fall in the same manner as if both beta factors were reduced. In the alternative regime, where it is optimal to tune onto one emitter, a difference in beta factors introduces an asymmetry in the behaviour of the average concurrence and it becomes preferable to tune onto the emitter that is more efficiently coupled to the waveguide. The two-photon entanglement generation process outperforms the single-photon process for all values of $\beta$. In Fig.~\ref{fig:fig3}(d) we show how two-photon input states can achieve a significant increase in $\mathcal{C}_\text{avg}$ over single photons for larger detunings.

In certain regimes the two-photon process is capable of generating maximal entanglement between spectrally distinct emitters where the single-photon process fails. We studied the robustness of this effect with respect to the system parameters. In Figs. \ref{fig:fig4}(a) and \ref{fig:fig4}(b) we show the maximum $\mathcal{C}_\text{avg}$ (optimised over photon frequency $\omega_{\rm opt}$) for the one- and two-photon input states as a function of the emitter detuning and the emitter line-width ratio. The single-photon case outperforms the two-photon case if the emitters are spectrally collocated, or if one of the emitters significant overlaps the other, however narrow the line-width. Crucially, however, by exploiting the multi-photon additivity of the phase shift, a two-photon process can efficiently generate entanglement for \emph{any} finite detuning without requiring arbitrarily small emitter lifetimes. The converse of this is also true: for any combination of line-widths $ \Gamma_{\text{1}}$ and $\Gamma_{\text{2}}$ there exists a non-zero emitter detuning which creates deterministic maximal entanglement given the optimal two-photon input state. In practice, this means a much greater freedom in matching solid-state emitters for entanglement generation in a Mach-Zehnder interferometer than previously thought.  
 
We extended the entanglement generation process to monochromatic $\ket{n,m}$ Fock states into the interferometer. In Figs.~\ref{fig:fig4}(c) and \ref{fig:fig4}(d) we show the maximum $\mathcal{C}_\text{avg}$ as a function of the emitter detuning and the emitter line-width ratio for input states $\ket{2,1}$ and $\ket{2,2}$, respectively (for more examples, see the Supplementary Information). There is a marked improvement in the entanglement generation over the single- and two-photon processes, with larger areas of parameter space achieving a near-unity $\mathcal{C}_\text{avg}$. Remarkably, this indicates that a wide range of imperfections in the fabrication of two identical emitters can be overcome by optical state optimisation. Note that the $\ket{2,1}$ case inherits features from both the $\ket{1,1}$ and $\ket{1,0}$ processes. It therefore performs well for both spectrally collocated emitters and those with finite detuning. A similar compound structure is visible in Fig.~\ref{fig:fig4}(d), where an input state $\ket{2,2}$ shows a double two-photon structure compared to the $\ket{1,1}$ input in Fig.~\ref{fig:fig4}(b). A clear trend emerges, where larger spectral emitter detuning can be overcome by higher number input states $\ket{n,m}$ (see Supplementary Information).

\begin{figure}[t]
\begin{center}
\includegraphics[width=8.5cm]{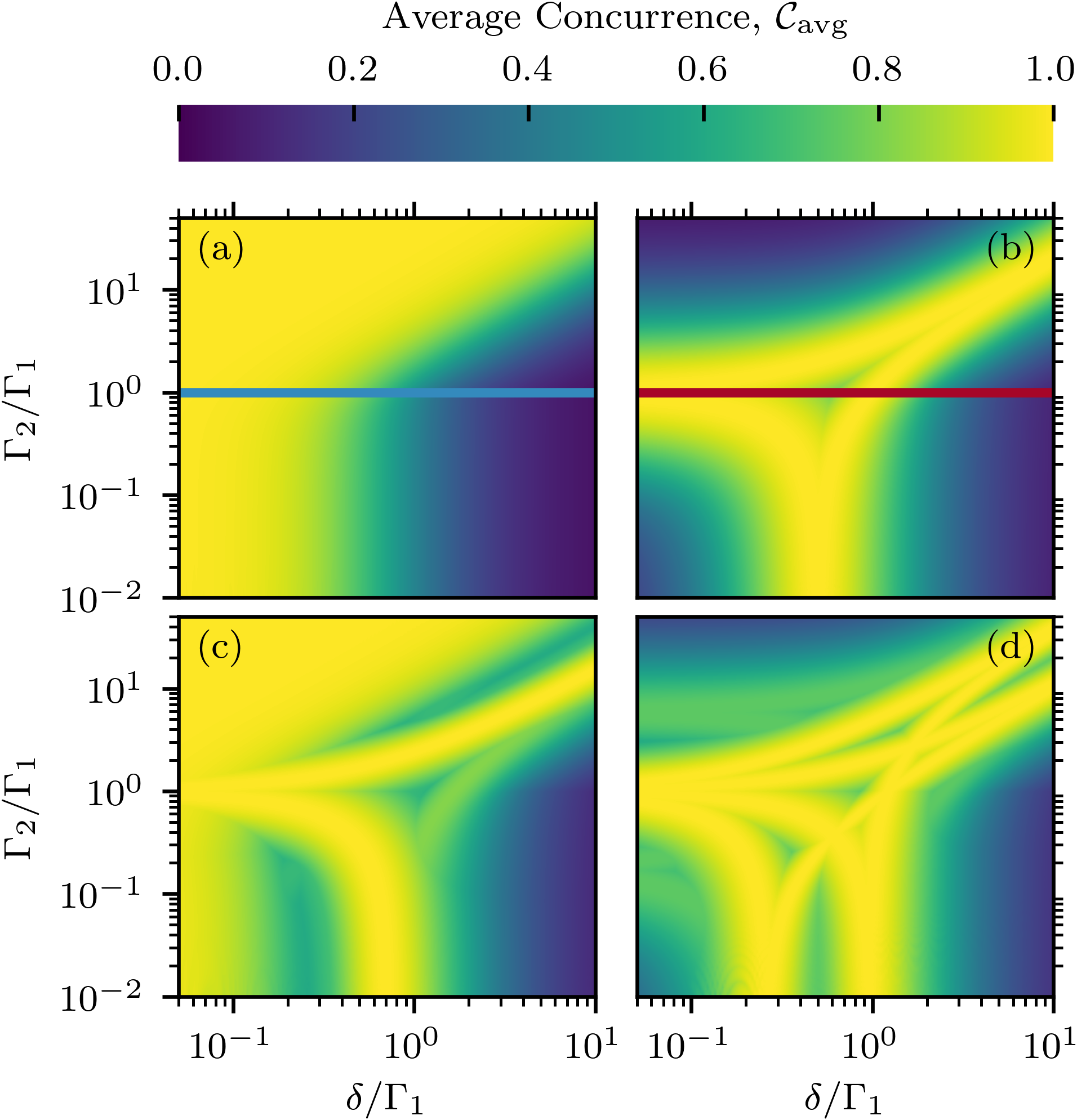}
\caption{Maximum average concurrence for different photon number input states injected into the interferometer. The emitter detuning $\delta$ and the line-width $\Gamma_2$ are both normalised to $\Gamma_1$, and we consider lossless waveguides ($\beta=1$). The input photons are identical and monochromatic in the configurations (a) $\ket{1,0}$; (b) $\ket{1,1}$; (c) $\ket{2,1}$; and (d) $\ket{2,2}$. The characteristic shapes in (a) and (b) recur in (c) and (d), and are also found in higher photon number input states $\ket{n,m}$.
The blue and red lines in (a) and (b) correspond respectively to the solid blue and red lines in Fig.~\ref{fig:fig3}(d).}
\label{fig:fig4}
\end{center}
\end{figure}

There are several practical challenges turning this entanglement generation process into a useful quantum technology. First, a reduced coupling of the emitter to the waveguide mode will reduce the phase shift imparted on the photons, and therefore lower the average concurrence. Second, the photons may be scattered out of the waveguide mode or be lost in the detection process. However, the use of fast, high-efficiency photon number resolving detectors will mitigate this problem, and such detectors are actively developed~\cite{Fiore08,Heath15}. Third, the photons must be created in tuneable identical quasi-monochromatic modes. There are a number of ways this can be achieved over a wide frequency range. Spontaneous Parametric Downconversion (SPDC) is inherently tuneable \cite{Klyshko70,Tanida12} and frequency filtering will create the optimal quasi-monochromatic pulses as well as remove unwanted frequency entanglement. The resulting photon generation rate reduction can be mitigated using multiplexing, which has been demonstrated both for SPDC photon sources \cite{Joshi18} and tuneable quantum dot sources \cite{Wang17}. Alternatively, tuning of single photon pulses is possible via frequency conversion \cite{Li16}. Finally, dephasing will have an impact on the entanglement generation process. The dominant dephasing mechanism for spin-doped solid-state emitters is nuclear spin interactions~\cite{Merkulov02}. While this naturally leads to a random precession of the spin ground-state, there are a number of strategies based on dynamical decoupling that may be used to suppress its impact~\cite{Viola99,Press10,Kuhlmann13}. In addition, solid-state emitters are subject to  charge fluctuations and phonon interactions. The former leads to spectral wandering occurring on a microsecond timescale, which may be overcome by operating the process on shorter timescales~\cite{Berthelot06}. Phonon scattering leads to sidebands~\cite{IlesSmith17b} that can be removed through frequency filtering or by placing the emitter in an optical cavity~\cite{Grange17,IlesSmith17a}.

In conclusion, we presented a robust entanglement generation mechanism between two solid-state qubits embedded in a Mach-Zehnder interferometer. Entangling techniques that use solid-state emitters are well-known to place very stringent requirements on the spectral identity of the emitters~\cite{Barrett05}. Our approach overcomes these restrictions by showing how to tailor multi-photon input states, mitigating a long considered weakness of solid-state emitters. We found that maximal deterministic entanglement between increasingly distinct emitters is possible using higher photon number input states, revealing a rich structure in multi-photon scattering from two emitters with different energies and line-widths. 
Our work provides a new methodology for solid-state entanglement generation, where the requirement for perfectly matched emitters can be relaxed in favour of optical state optimisation.

\paragraph{Acknowledgements.---}  the authors thank A.P. Foster, D. Hallett, and M.S. Skolnick for useful suggestions, and EPSRC for financial support. K.B.J and J.M. acknowledge funding from the Danish Council for Independent Research (DFF-4181-00416). 
J.I.S. is supported by the Royal Commission for the Exhibition of 1851.

\noindent $^\dagger$\,DLH and KBJ contributed equally to this project.


%

\end{document}


\title{Supplemental Information: Generating maximal entanglement between spectrally distinct emitters}
\author{David L. Hurst}
\affiliation{Department of Physics and Astronomy, University of Sheffield, Hounsfield Road, Sheffield, S3 7RH, United Kingdom}
\author{Kristoffer B. Joanesarson}
\affiliation{Department of Physics and Astronomy, University of Sheffield, Hounsfield Road, Sheffield, S3 7RH, United Kingdom}
\affiliation{Department of Photonics Engineering, Technical University of Denmark, \O rsteds Plads, Lyngby, Denmark}
\author{Jake Iles-Smith}
\affiliation{Department of Physics and Astronomy, University of Sheffield, Hounsfield Road, Sheffield, S3 7RH, United Kingdom}
\author{Jesper M\o rk}
\affiliation{Department of Photonics Engineering, Technical University of Denmark, \O rsteds Plads, Lyngby, Denmark}
\author{Pieter Kok}
\affiliation{Department of Physics and Astronomy, University of Sheffield, Hounsfield Road, Sheffield, S3 7RH, United Kingdom}
\date{\today}

\maketitle

\section{Figure of Merit}
\noindent
Concurrence is the entanglement measure we choose, as it permits a natural extension to states that can only be described in terms of a density matrix~\cite{PhysRevLett.80.2245}. For the two-qubit mixed state $ \rho $, the concurrence is found via
\begin{align}
\mathcal{C}\qty(\rho) = \mathrm{max}\qty(0,\lambda_1-\lambda_2-\lambda_3-\lambda_4), \label{eq:conc}
\end{align}
where $ \lambda_i$ is the $i$\textsuperscript{th} eigenvalue of
\begin{align}
R\qty(\rho) = \sqrt{\sqrt{\rho} \sigma^{(1)}_y \otimes \sigma^{(2)}_y  \rho^* \sigma^{(1)}_y \otimes \sigma^{(2)}_y \sqrt{\rho}},
\end{align}
arranged in decreasing order. For a pure two-qubit state $ \ket{\phi} $ the expression for the concurrence can be written as $ \mathcal{C} \qty(\ket{\phi})= |\mel{\phi}{\sigma_y^{(1)}\otimes\sigma_y^{(2)}}{\phi^*}| $. In the projection-by-measurement type protocol, the probability that detector $ 1 $ ($ D_1 $) and $ 2 $ ($ D_2 $) gives $ p $ and $ q $ clicks, respectively, is $ \mathrm{Pr}(p,q) $. The detection outcome heralds the qubit state $ \rho_{\qty(p,q)} $, and $ \mathcal{C}\qty(\rho_{\qty(p,q)})$ is the associated concurrence. The single set $ \qty(p,q) $ corresponds to one of the possible detection outcomes where the number of possible outcomes depends on the number of incident photons, whether the detectors are number resolving or not, and whether photon-loss is present.  We define the average concurrence as the figure of merit and this is given by

\begin{align}
\mathcal{C}_{\text{avg}} \equiv \sum_{\qty(p,q)} \mathrm{Pr}(p,q)\mathcal{C}\qty(\rho_{\qty(p,q)}).
\end{align}

\section{\label{sec:1phConc}Single-photon protocol}

\subsection{\label{ssec:1Photon_transport}Single-photon transport in the MZI}
\noindent
The initial two-qubit state is taken to be the separable pure state
\begin{align}
\ket{\phi_\text{in}} &= \frac{1}{2}\qty(\ket{\uparrow} +\ket{\downarrow})\otimes \qty(\ket{\uparrow} + \ket{\downarrow}).
\end{align}
We choose a single-photon input state with the photon injected into the upper arm of the Mach-Zehnder Interferometer (MZI). The optical state is
\begin{align}
\ket{\psi_\text{in}} = \int \dd{\omega}\, \xi(\omega) u^{\dagger}(\omega)\ket{\varnothing},
\end{align}
where $u(\omega)$ is the annihilation operator for a photon of frequency $\omega$ in the upper interferometer arm and $ \xi(\omega) $ describes the optical wavepacket, such that $ \int \dd{\omega}\qty|\xi(\omega)|^2 = 1 $. The photon state after scattering off the first beam-splitter (BS) is
\begin{align}
\ket{\psi_2} = \int \dd{\omega}\, \xi(\omega) \frac{1}{\sqrt{2}}\qty[u^{\dagger}(\omega) +  d^{\dagger}(\omega)]\ket{\varnothing},
\end{align}
where $d(\omega)$ is the annihilation operator for the lower interferometer arm. The combined photon-emitter state is 
\begin{align}
\ket{\Psi_2} &=\ket{\phi_\text{in}} \otimes \ket{\psi_2} = \frac{1}{2}\qty(\ket{\uparrow} +\ket{\downarrow})\otimes \qty(\ket{\uparrow} + \ket{\downarrow}) \otimes  \int \dd{\omega}\, \xi(\omega) \frac{1}{\sqrt{2}}\qty[u^{\dagger}(\omega) +  d^{\dagger}(\omega)]\ket{\varnothing}.
\end{align}
After scattering off the emitters we find that the state evolves to
\begin{align}
\begin{split}
\ket{\Psi_3} &= \frac{1}{2\sqrt{2}}\int \dd{\omega}\, \xi(\omega) \qty(t_1(\omega)\ket{\uparrow} +\ket{\downarrow})\otimes \qty(\ket{\uparrow} + \ket{\downarrow}) \otimes   u^{\dagger}(\omega)\ket{\varnothing}
\\
&\quad  +\frac{1}{2\sqrt{2}}\int \dd{\omega}\, \xi(\omega) \qty(\ket{\uparrow} +\ket{\downarrow})\otimes \qty(t_2(\omega)\ket{\uparrow} + \ket{\downarrow}) \otimes   d^{\dagger}(\omega)\ket{\varnothing},
\end{split}
\end{align}
where $t_{1}(\omega)$ and $t_{2}(\omega)$ are the transmission coefficients for a photon of frequency $\omega$ scattering from the emitter at position $1$ and $ 2 $, respectively (defined in the main text). The photon state after the second BS and before measurement is
\begin{align}
\begin{split}
\ket{\Psi_4} &= \frac{1}{4}\int \dd{\omega}\, \xi(\omega) \bigg(\qty[t_1(\omega)+t_2(\omega)]\ket{\uparrow, \uparrow} + \qty[t_1(\omega)+1] \ket{\uparrow, \downarrow}  + \qty[1+t_2(\omega)]\ket{\downarrow, \uparrow} +2 \ket{\downarrow ,\downarrow}\bigg) \otimes   u^{\dagger}(\omega)\ket{\varnothing}
\\
&\quad  + \frac{1}{4}\int \dd{\omega}\, \xi(\omega) \bigg(\qty[-t_1(\omega)+t_2(\omega)]\ket{\uparrow ,\uparrow} + \qty[-t_1(\omega)+1]\ket{\uparrow, \downarrow}  + \qty[-1+t_2(\omega)]\ket{\downarrow, \uparrow} \bigg) \otimes   d^{\dagger}(\omega)\ket{\varnothing} \label{eq:1ph_Psi4}
\end{split}
\end{align}

\subsection{\label{ssec:1Photon_measurement}Single-photon measurement}
\noindent
We define our positive operator-valued measures (POVMs) used to determine the detection probabilities and post measurement states.
The single-photon POVMs are
\begin{subequations}
\begin{align}
\Pi(u) &= \int \dd{\omega} \, u^{\dagger}(\omega) \ketbra{\varnothing}u(\omega) ,
\\
\Pi(d) &= \int \dd{\omega} \, d^{\dagger}(\omega) \ketbra{\varnothing}d(\omega) .
\end{align}
\end{subequations}
The detection probabilities associated with the POVMs and the corresponding post-measurement states are given by \cite{nielsen2002quantum}
\begin{align}
\mathrm{Pr}\qty(p,q) = \mel{\Psi_4}{\Pi(u^p d^q)}{\Psi_4}, \quad \text{and} \quad \ket{\Psi'_{\qty(p,q)}}=\frac{1}{\sqrt{\mathrm{Pr}\qty(p,q)}}\Pi(u^p d^q)\ket{\Psi_4},\label{eq:Prob_postState}
\end{align}
respectively. Using the state in Eq. (\ref{eq:1ph_Psi4}), the probability of detecting one photon at $ D_1 $ is
\begin{align}
\mathrm{Pr}(1,0) = \frac{1}{16}\int \dd{\omega}\, |\xi(\omega)|^2\qty(|t_1(\omega)+t_2(\omega)|^2 + |t_1(\omega)+1|^2 + |1+t_2(\omega)|^2 + 4),
\end{align}
and at $ D_2 $ is
\begin{align}
\mathrm{Pr}(0,1) = \frac{1}{16}\int \dd{\omega}\, |\xi(\omega)|^2\qty(|-t_1(\omega)+t_2(\omega)|^2 + |-t_1(\omega)+1|^2 + |-1+t_2(\omega)|^2 ).
\end{align}
We find the two-qubit state by tracing out the waveguide modes. This gives us the density matrices
\begin{align}
\begin{split}
\rho_{\qty(1,0)} &= \int \dd{\omega}\, \qty|\xi(\omega)|^2 \qty[\ketbra{\phi_{\qty(1,0)}(\omega)}],
\end{split}
\end{align}
and
\begin{align}
\begin{split}
\rho_{\qty(0,1)} &= \int \dd{\omega}\, \qty|\xi(\omega)|^2 \qty[\ketbra{\phi_{\qty(0,1)}(\omega)}],
\end{split}
\end{align}
where
\begin{align}
\begin{split}
\ket{\phi_{\qty(1,0)}(\omega)} &= \frac{1}{4\sqrt{\mathrm{Pr}(1,0)}} \bigg[ \qty[t_1(\omega)+t_2(\omega)]\ket{\uparrow ,\uparrow} + \qty[t_1(\omega)+1] \ket{\uparrow ,\downarrow}  + \qty[1+t_2(\omega)]\ket{\downarrow ,\uparrow} +2 \ket{\downarrow ,\downarrow} \bigg],
\end{split}
\end{align}
and
\begin{align}
\begin{split}
\ket{\phi_{\qty(0,1)}(\omega)} &= \frac{1}{4\sqrt{\mathrm{Pr}(0,1)}} \bigg[ \qty[-t_1(\omega)+t_2(\omega)]\ket{\uparrow, \uparrow} + \qty[-t_1(\omega)+1] \ket{\uparrow, \downarrow}  + \qty[-1+t_2(\omega)]\ket{\downarrow, \uparrow}  \bigg].
\end{split}
\end{align}
It is from these density matrices we determine the concurrence according to Eq.~\eqref{eq:conc}.

\subsection{\label{ssec:1Photon_conc_mono}Monochromatic single-photon input}
\noindent
For a monochromatic input-photon with frequency $ \omega $, the two density operators describing the heralded two-qubit states can be reduced to pure states. For the probabilities and corresponding two-qubit states we find the following simplified results.
With a probability of
\begin{align}
\mathrm{Pr}(1,0) &= \frac{1}{16}\qty(|t_1(\omega)+t_2(\omega)|^2 + |t_1(\omega)+1|^2 + |1+t_2(\omega)|^2 + 4),
\end{align}
we detect a photon at $D_1$ and thus herald the two-qubit state
\begin{align}
\ket{\phi_{\qty(1,0)}(\omega)} &= \frac{\qty[t_1(\omega)+t_2(\omega)]\ket{\uparrow ,\uparrow} + \qty[t_1(\omega)+1] \ket{\uparrow ,\downarrow}  + \qty[1+t_2(\omega)]\ket{\downarrow, \uparrow} +2 \ket{\downarrow, \downarrow}}{\sqrt{|t_1(\omega)+t_2(\omega)|^2 + |t_1(\omega)+1|^2 + |1+t_2(\omega)|^2 + 4}}.
\end{align}
This state has a concurrence of
\begin{align}
\mathcal{C}\qty(\ket{\phi_{\qty(1,0)}(\omega)}) = \frac{\qty|4\qty[t_1(\omega)+t_2(\omega)] -2\qty[t_1(\omega)+1]\qty[1+t_2(\omega)] |}{|t_1(\omega)+t_2(\omega)|^2 + |t_1(\omega)+1|^2 + |1+t_2(\omega)|^2 + 4}.
\end{align}
Similarly, we detect a photon at $D_2$ with probability
\begin{align}
\mathrm{Pr}(0,1) &= \frac{1}{16}\qty(|-t_1(\omega)+t_2(\omega)|^2 + |-t_1(\omega)+1|^2 + |-1+t_2(\omega)|^2 ),
\end{align}
which heralds the two-qubit state
\begin{align}
\ket{\phi_{\qty(0,1)}(\omega)} &= \frac{\qty[-t_1(\omega)+t_2(\omega)]\ket{\uparrow ,\uparrow} + \qty[-t_1(\omega)+1] \ket{\uparrow, \downarrow}  + \qty[-1+t_2(\omega)]\ket{\downarrow ,\uparrow} }{\sqrt{|-t_1(\omega)+t_2(\omega)|^2 + |-t_1(\omega)+1|^2 + |-1+t_2(\omega)|^2 }},
\end{align}
with the corresponding concurrence
\begin{align}
\mathcal{C}\qty(\ket{\phi_{\qty(0,1)}(\omega)}) = \frac{2\qty|\qty[-t_1(\omega)+1]\qty[-1+t_2(\omega)]|}{|-t_1(\omega)+t_2(\omega)|^2 + |-t_1(\omega)+1|^2 + |-1+t_2(\omega)|^2}.
\end{align}

The average concurrence in this case would be
\begin{align}
\mathcal{C}_{\mathrm{avg}}(\omega) = \mathrm{Pr}(1,0)\mathcal{C}\qty(\ket{\phi_{\qty(1,0)}(\omega)}) + \mathrm{Pr}(0,1)\mathcal{C}\qty(\ket{\phi_{\qty(0,1)}(\omega)}).
\end{align}

\section{\label{sec:2phConc}Two-photon protocol}

\subsection{\label{ssec:2Photon_transport}Two-photon transport in the MZI}
\noindent
The two-photon input state is
\begin{align}
\ket{\psi_1} = \int \dd{\omega}\dd{\omega'}\, \xi(\omega,\omega') u^{\dagger}(\omega)d^{\dagger}(\omega')\ket{\varnothing},
\end{align}
normalized such that $ \int  \dd{\omega}\dd{\omega'}\, \qty|\xi(\omega,\omega')|^2 = 1 $. After the first BS, the optical state is
\begin{align}
\begin{split}
\ket{\psi_2} &= \frac{1}{2}\int \dd{\omega}\dd{\omega'}\,  \bigg[-\xi(\omega,\omega')u^{\dagger}(\omega)u^{\dagger}(\omega') + \xi(\omega,\omega')d^{\dagger}(\omega)d^{\dagger}(\omega')   + \delta \xi(\omega,\omega')u^{\dagger}(\omega)d^{\dagger}(\omega')\bigg]\ket{\varnothing},
\end{split}
\end{align}
where we have defined $\delta \xi(\omega,\omega') =  \xi(\omega,\omega')-\xi(\omega',\omega) $. We see that in the case of indistinguishable photons the last term cancels out, due to the well-known Hong-Ou-Mandel (HOM) interference effect \cite{PhysRevLett.59.2044}. The combined photon-emitter state is again
\begin{align}
\ket{\Psi_2} &=\ket{\phi_\text{in}} \otimes \ket{\psi_2}.
\end{align}
After scattering off the emitters we find
\begin{align}
\begin{split}
\ket{\Psi_3} &= \frac{1}{4}\int \dd{\omega}\dd{\omega'}\, \qty(-\tilde{\xi}_1(\omega,\omega') \ket{\uparrow} -\xi(\omega,\omega')\ket{\downarrow})\otimes \qty(\ket{\uparrow} + \ket{\downarrow}) \otimes   u^{\dagger}(\omega)u^{\dagger}(\omega')\ket{\varnothing}
\\
&\quad  +\frac{1}{4}\int \dd{\omega}\dd{\omega'}\, \qty(\ket{\uparrow} + \ket{\downarrow})\otimes \qty(\tilde{\xi}_2(\omega,\omega')\ket{\uparrow} + \xi(\omega,\omega') \ket{\downarrow}) \otimes   d^{\dagger}(\omega)d^{\dagger}(\omega')\ket{\varnothing}
\\
&\quad  +\frac{1}{4}\int \dd{\omega}\dd{\omega'}\,\delta \xi(\omega,\omega') \qty(t_1(\omega)\ket{\uparrow} + \ket{\downarrow})\otimes \qty(t_2(\omega)\ket{\uparrow} +  \ket{\downarrow}) \otimes   u^{\dagger}(\omega)d^{\dagger}(\omega')\ket{\varnothing}. \label{eq:2ph_Psi3_noLoss}
\end{split}
\end{align}
We have introduced two new envelopes, $ \tilde{\xi}_1 $ and $ \tilde{\xi}_2 $, which for finite widths are not simple products of the initial envelope and a scattering coefficient. The envelopes can be written as a sum of linear interactions and an additional bound state term. These envelopes can be found in e.g. Ref.~\cite{PhysRevA.91.063823}
\begin{align}
\begin{split}
\tilde{\xi}_{\alpha}(\omega,\omega') &= \frac{1}{2}t_{\alpha}(\omega)t_{\alpha}(\omega')\qty[\xi(\omega,\omega') + \xi(\omega',\omega)]
\\
& \qquad + \frac{i}{2}\frac{\sqrt{\Gamma_{\alpha}}}{\pi}s_{\alpha}(\omega)s_{\alpha}(\omega') \int \dd{k}\, \qty[s_{\alpha}(k)+s_{\alpha}(\omega+\omega'-k)]\xi(k,\omega + \omega'-k),
\end{split}
\end{align}
with 
\begin{align}
s_{\alpha}(\omega) = \frac{\sqrt{\Gamma_{\alpha}}}{\omega-E_{\alpha}/\hbar + i\qty(\Gamma_{\alpha}+\gamma_{\alpha})/2}.
\end{align}
If photon-loss is present the results will depend on whether or not the detectors are assumed to number resolving or not. We therefore introduce two additional channels, one leakage channel for each emitter, and modify Eq.~(\ref{eq:2ph_Psi3_noLoss}) to
\begin{align}
\begin{split}
\ket{\Psi_3} &= \frac{1}{4}\int \dd{\omega}\dd{\omega'}\, \qty(-\tilde{\xi}_1(\omega,\omega') \ket{\uparrow} -\xi(\omega,\omega')\ket{\downarrow})\otimes \qty(\ket{\uparrow} + \ket{\downarrow}) \otimes   u^{\dagger}(\omega)u^{\dagger}(\omega')\ket{\varnothing}
\\
&\quad  +\frac{1}{4}\int \dd{\omega}\dd{\omega'}\, \qty(\ket{\uparrow} + \ket{\downarrow})\otimes \qty(\tilde{\xi}_2(\omega,\omega')\ket{\uparrow} + \xi(\omega,\omega') \ket{\downarrow}) \otimes   d^{\dagger}(\omega)d^{\dagger}(\omega')\ket{\varnothing}
\\
&\quad  +\frac{1}{4}\int \dd{\omega}\dd{\omega'}\,\delta \xi(\omega,\omega') \qty(t_1(\omega)\ket{\uparrow} + \ket{\downarrow})\otimes \qty(t_2(\omega')\ket{\uparrow} +  \ket{\downarrow}) \otimes   u^{\dagger}(\omega)d^{\dagger}(\omega')\ket{\varnothing}
\\
&\quad  +\frac{1}{4}\int \dd{\omega}\dd{\omega'}\, \qty(-\tilde{\xi}_1^{(r)}(\omega,\omega')\ket{\uparrow} +0 \ket{\downarrow})\otimes \qty(\ket{\uparrow} +  \ket{\downarrow}) \otimes   u^{\dagger}(\omega)r_1^{\dagger}(\omega')\ket{\varnothing}
\\
&\quad  +\frac{1}{4}\int \dd{\omega}\dd{\omega'}\,\delta \xi(\omega,\omega') \qty(t_1(\omega)\ket{\uparrow} + \ket{\downarrow})\otimes \qty(t_2^{(r)}(\omega')\ket{\uparrow} + 0 \ket{\downarrow})\otimes   u^{\dagger}(\omega)r_2^{\dagger}(\omega')\ket{\varnothing}
\\
&\quad  +\frac{1}{4}\int \dd{\omega}\dd{\omega'}\,\delta \xi(\omega,\omega') \qty(t_1^{(r)}(\omega)\ket{\uparrow} +0 \ket{\downarrow})\otimes \qty(t_2(\omega')\ket{\uparrow} +  \ket{\downarrow}) \otimes   r_1^{\dagger}(\omega)d^{\dagger}(\omega')\ket{\varnothing}
\\
&\quad  +\frac{1}{4}\int \dd{\omega}\dd{\omega'}\,\qty(\ket{\uparrow} + \ket{\downarrow})\otimes \qty(\tilde{\xi}_2^{(r)}(\omega,\omega')\ket{\uparrow} +  0\ket{\downarrow}) \otimes   d^{\dagger}(\omega)r_2^{\dagger}(\omega')\ket{\varnothing}
\\
&\quad  +\frac{1}{4}\int \dd{\omega}\dd{\omega'}\,  \qty(-\tilde{\xi}_1^{(rr)}(\omega,\omega')\ket{\uparrow} - 0 \ket{\downarrow})\otimes \qty(\ket{\uparrow} +  \ket{\downarrow}) \otimes   r_1^{\dagger}(\omega)r_1^{\dagger}(\omega')\ket{\varnothing}
\\
&\quad  +\frac{1}{4}\int \dd{\omega}\dd{\omega'}\,\delta \xi(\omega,\omega') \qty(t_1^{(r)}(\omega)\ket{\uparrow} +0 \ket{\downarrow})\otimes \qty(t_2^{(r)}(\omega')\ket{\uparrow} + 0 \ket{\downarrow}) \otimes   r_1^{\dagger}(\omega)r_2^{\dagger}(\omega')\ket{\varnothing}
\\
&\quad  +\frac{1}{4}\int \dd{\omega}\dd{\omega'}\, \qty(\ket{\uparrow} + \ket{\downarrow})\otimes \qty(\tilde{\xi}_2^{(rr)}(\omega,\omega')\ket{\uparrow} + 0 \ket{\downarrow}) \otimes   r_2^{\dagger}(\omega)r_2^{\dagger}(\omega')\ket{\varnothing},
\end{split} \label{eq:big}
\end{align}
where $r_1$ and $r_2$ are annihilation operators for photons in the reservoirs around the emitters at positions $1$ and $2$, respectively. The newly introduced reservoir scattering coefficients and envelopes are
\begin{align}
t_{\alpha}^{(r)}(\omega) &= \frac{-i\sqrt{\Gamma_{\alpha}\gamma_{\alpha}}}{\omega-E_{\alpha}/\hbar + i\qty(\Gamma_{\alpha}+\gamma_{\alpha})/2},
\end{align}
and
\begin{align}
\begin{split}
\tilde{\xi}_{\alpha}^{(r)}(\omega,\omega') &= t_{\alpha}(\omega)t^{(r)}_{\alpha}(\omega')\xi(\omega,\omega') +t^{(r)}_{\alpha}(\omega)t_{\alpha}(\omega')\xi(\omega',\omega)
\\
& \qquad + \frac{\sqrt{\Gamma_{\alpha}}}{\pi}s_{\alpha}(\omega)s^{(r)}_{\alpha}(\omega') \int \dd{k}\, \qty[s_{\alpha}(k)+s_{\alpha}(\omega+\omega'-k)]\xi(k,\omega + \omega'-k),
\end{split}
\\
\begin{split}
\tilde{\xi}_{\alpha}^{(rr)}(\omega,\omega') &= \frac{1}{2}t^{(r)}_{\alpha}(\omega)t^{(r)}_{\alpha}(\omega')\qty[\xi(\omega,\omega') + \xi(\omega',\omega)]
\\
& \qquad + \frac{i}{2}\frac{\sqrt{\Gamma_{\alpha}}}{\pi}s^{(r)}_{\alpha}(\omega)s^{(r)}_{\alpha}(\omega') \int \dd{k}\, \qty[s_{\alpha}(k)+s_{\alpha}(\omega+\omega'-k)]\xi(k,\omega + \omega'-k),
\end{split}
\end{align}
respectively, where we have defined
\begin{align}
s^{(r)}_{\alpha}(\omega) = \frac{\sqrt{\gamma_{\alpha}}}{\omega-E_{\alpha}/\hbar + i\qty(\Gamma_{\alpha}+\gamma_{\alpha})/2}.
\end{align}

For notational convenience we condense Eq.~\eqref{eq:big} to
\begin{align}
\begin{split}
\ket{\Psi_3} &\equiv\int \dd{\omega}\dd{\omega'} \bigg[ \ket{\phi^{(3)}_{uu}(\omega,\omega')}\otimes u^{\dagger}(\omega)u^{\dagger}(\omega') + \ket{\phi^{(3)}_{dd}(\omega,\omega')}\otimes d^{\dagger}(\omega)d^{\dagger}(\omega')
\\
& \quad  + \ket{\phi^{(3)}_{ud}(\omega,\omega')}\otimes u^{\dagger}(\omega)d^{\dagger}(\omega') + \ket{\phi^{(3)}_{ur_1}(\omega,\omega')}\otimes u^{\dagger}(\omega)r_1^{\dagger}(\omega')
\\
& \quad + \ket{\phi^{(3)}_{ur_2}(\omega,\omega')}\otimes u^{\dagger}(\omega)r_2^{\dagger}(\omega') + \ket{\phi^{(3)}_{r_1d}(\omega,\omega')}\otimes r_1^{\dagger}(\omega)d^{\dagger}(\omega')
\\
& \quad +\ket{\phi^{(3)}_{dr_2}(\omega,\omega')}\otimes d^{\dagger}(\omega)r_2^{\dagger}(\omega') + \ket{\phi_{r_1r_1}(\omega,\omega')}\otimes r_1^{\dagger}(\omega)r_1^{\dagger}(\omega')
\\
& \quad +\ket{\phi_{r_1r_2}(\omega,\omega')}\otimes r_1^{\dagger}(\omega)r_2^{\dagger}(\omega') + \ket{\phi_{r_2r_2}(\omega,\omega')}\otimes r_2^{\dagger}(\omega)r_2^{\dagger}(\omega')\bigg]\ket{\varnothing}, \label{eq:2Ph_Psi3}
\end{split}
\end{align}
and evolve it through the second BS so that
\begin{align}
\begin{split}
\ket{\Psi_4} &= \int \dd{\omega}\dd{\omega'} \bigg[ \ket{\phi_{uu}(\omega,\omega')}\otimes u^{\dagger}(\omega)u^{\dagger}(\omega') + \ket{\phi_{dd}(\omega,\omega')}\otimes d^{\dagger}(\omega)d^{\dagger}(\omega')
\\
& \quad  + \ket{\phi_{ud}(\omega,\omega')}\otimes u^{\dagger}(\omega)d^{\dagger}(\omega') + \ket{\phi_{ur_1}(\omega,\omega')}\otimes u^{\dagger}(\omega)r_1^{\dagger}(\omega')
\\
& \quad + \ket{\phi_{ur_2}(\omega,\omega')}\otimes u^{\dagger}(\omega)r_2^{\dagger}(\omega') + \ket{\phi_{r_1d}(\omega,\omega')}\otimes r_1^{\dagger}(\omega)d^{\dagger}(\omega')
\\
& \quad +\ket{\phi_{dr_2}(\omega,\omega')}\otimes d^{\dagger}(\omega)r_2^{\dagger}(\omega') + \ket{\phi_{r_1r_1}(\omega,\omega')}\otimes r_1^{\dagger}(\omega)r_1^{\dagger}(\omega')
\\
& \quad +\ket{\phi_{r_1r_2}(\omega,\omega')}\otimes r_1^{\dagger}(\omega)r_2^{\dagger}(\omega') + \ket{\phi_{r_2r_2}(\omega,\omega')}\otimes r_2^{\dagger}(\omega)r_2^{\dagger}(\omega')\bigg]\ket{\varnothing},
\end{split}
\end{align}
where
\begin{subequations}
\begin{align}
\ket{\phi_{uu}(\omega,\omega')} &= \frac{1}{2}\qty(\ket{\phi^{(3)}_{uu}(\omega,\omega')} + \ket{\phi^{(3)}_{d d}(\omega,\omega')} + \ket{\phi^{(3)}_{u d}(\omega,\omega')}),
\\
\ket{\phi_{dd}(\omega,\omega')} &= \frac{1}{2}\qty(\ket{\phi^{(3)}_{uu}(\omega,\omega')} + \ket{\phi^{(3)}_{d d}(\omega,\omega')} - \ket{\phi^{(3)}_{u d}(\omega,\omega')}),
\\
\ket{\phi_{ud}(\omega,\omega')} &= -\ket{\phi^{(3)}_{uu}(\omega,\omega')} + \ket{\phi^{(3)}_{d d}(\omega,\omega')} +\frac{1}{2}\qty( \ket{\phi^{(3)}_{u d}(\omega,\omega')} -  \ket{\phi^{(3)}_{u d}(\omega',\omega)}),
\\
\ket{\phi_{ur_1}(\omega,\omega')} &= \frac{1}{\sqrt{2}}\qty(\ket{\phi^{(3)}_{ur_1}(\omega,\omega')} + \ket{\phi^{(3)}_{r_1 d}(\omega',\omega)}),
\\
\ket{\phi_{ur_2}(\omega,\omega')} &= \frac{1}{\sqrt{2}}\qty(\ket{\phi^{(3)}_{ur_2}(\omega,\omega')} + \ket{\phi^{(3)}_{dr_2}(\omega',\omega)}),
\\
\ket{\phi_{r_1d}(\omega,\omega')} &= \frac{1}{\sqrt{2}}\qty(-\ket{\phi^{(3)}_{ur_1}(\omega',\omega)} + \ket{\phi^{(3)}_{r_1 d}(\omega,\omega')}),
\\
\ket{\phi_{d r_2}(\omega,\omega')} &= \frac{1}{\sqrt{2}}\qty(-\ket{\phi^{(3)}_{ur_2}(\omega,\omega')} + \ket{\phi^{(3)}_{d r_2}(\omega,\omega')}).
\end{align}
\end{subequations}

\subsection{\label{ssec:2Photon_measurement}Two-photon measurement}
\noindent
We define a set of two-photon POVMs to determine the detection probabilities and post-measurement states. We use $ \mu $ ($ \mu^{\dagger} $) and $ \nu $ ($ \nu^{\dagger} $) to denote the pre-defined bosonic annihilation (creation) operators. The POVMs are
\begin{align}
\Pi(\mu ,\nu) &= f_{\mu \nu} \int \dd{\omega}\dd{\omega'} \, \mu^{\dagger}(\omega)\nu^{\dagger}(\omega') \ketbra{\varnothing} \mu(\omega)\nu(\omega'), \qquad \text{where} \qquad f_{\mu \nu} = \begin{cases}
\frac{1}{2} & \text{if}\quad \mu=\nu, \\
1 & \text{if}\quad \mu\neq\nu. \\
\end{cases}
\end{align}
The detection probabilities associated with the POVMs and corresponding post-measurement states are determined from \eqref{eq:Prob_postState}. The heralded two-qubit density matrices are found from tracing out the fields.

In the absence of any loss we need only to concern our selves with the probabilities 
\begin{align}
\mathrm{Pr}\qty(2,0) &= \mel{\Psi_4}{\Pi(u, u)}{\Psi_4}, \qquad 
\mathrm{Pr}\qty(0,2) = \mel{\Psi_4}{\Pi(d, d)}{\Psi_4}, \qquad \mathrm{Pr}\qty(1,1) = \mel{\Psi_4}{\Pi(u, d)}{\Psi_4},
\end{align}
and the respective heralded density matrices
\begin{subequations}
\begin{align}
\rho_{\qty(2,0)} &= \frac{1}{\mathrm{Pr}(2,0)} \mathrm{Tr}_{\mathrm{field}} \qty{\Pi(u, u)\ketbra{\Psi_4}\Pi(u,u) } ,
\\
\rho_{\qty(0,2)} &= \frac{1}{\mathrm{Pr}(0,2)} \mathrm{Tr}_{\mathrm{field}} \qty{\Pi(d,d)\ketbra{\Psi_4}\Pi(d,d) } ,
\\
\rho_{\qty(1,1)} &= \frac{1}{\mathrm{Pr}(1,1)} \mathrm{Tr}_{\mathrm{field}} \qty{\Pi(u, d)\ketbra{\Psi_4}\Pi(u,d) } .
\end{align}
\end{subequations}
All other terms will disappear. In that case we do not need photon-number resolving detectors since we can infer a two-photon detection even when we do not get a coincidence click. If we include dissipation, considerations regarding the detector type is required.

\subsubsection{Photon-number resolving detectors}
\noindent
In case where we have access to photon-number resolving detectors, we can distinguish between detecting a two-photon state in one of the detectors and detecting only one photon in the same detector. We cannot, however, know whether we lost a photon from via interaction with emitter 1 or 2. The relevant probabilities are therefore
\begin{subequations}
\begin{align}
\mathrm{Pr}\qty(1,0) &= \mel{\Psi_4}{\Pi(u, r_1)}{\Psi_4} + \mel{\Psi_4}{\Pi(u, r_2)}{\Psi_4},
\\
\mathrm{Pr}\qty(0,1) &= \mel{\Psi_4}{\Pi(r_1, d)}{\Psi_4} + \mel{\Psi_4}{\Pi(d, r_2)}{\Psi_4},
\\
\mathrm{Pr}\qty(0,0) &= \mel{\Psi_4}{\Pi(r_1, r_1)}{\Psi_4} + \mel{\Psi_4}{\Pi(r_1, r_2)}{\Psi_4} + \mel{\Psi_4}{\Pi(r_2, r_2)}{\Psi_4},
\end{align}
\end{subequations}
and the corresponding density matrices are
\begin{subequations}
\begin{align}
\rho_{\qty(1,0)} &= \frac{1}{\mathrm{Pr}(1,0)} \mathrm{Tr}_{\mathrm{field}} \qty{\qty(\Pi(u, r_1)+\Pi(u, r_2))\ketbra{\Psi_4}\qty(\Pi(u, r_1)+\Pi(u, r_2)) } ,
\\
\rho_{\qty(0,1)} &= \frac{1}{\mathrm{Pr}(0,1)} \mathrm{Tr}_{\mathrm{field}} \qty{\qty(\Pi(r_1,d)+\Pi(d,r_2))\ketbra{\Psi_4}\qty(\Pi(r_1,d)+\Pi(d,r_2)) } ,
\\
\rho_{\qty(0,0)} &= \frac{1}{\mathrm{Pr}(0,0)} \mathrm{Tr}_{\mathrm{field}} \qty{\qty(\Pi(r_1,r_1)+\Pi(r_1,r_2)+\Pi(r_2,r_2))\ketbra{\Psi_4}\qty(\Pi(r_1,r_1)+\Pi(r_1,r_2)+\Pi(r_2,r_2) ) } .
\end{align}
\end{subequations}
We can use these density matrices to determine the corresponding concurrences using Eq.~\eqref{eq:conc}.

\subsubsection{Non-photon-number resolving detectors}
\noindent
If photon-number resolving detectors is not assumed, we cannot distinguish between having lost a single photon and detecting a two-photon state on a single detector. We can, however, still infer the state from a coincidence detection or no detection at all. There are therefore only four possible detection outcomes with the probability given by
\begin{subequations}
\begin{align}
\mathrm{Pr'}(1,1) &=  \mathrm{Pr}(1,1),
\\
\mathrm{Pr'}(1,0) &=  \mathrm{Pr}(2,0) + \mathrm{Pr}(1,0),
\\
\mathrm{Pr'}(0,1) &=  \mathrm{Pr}(0,2) + \mathrm{Pr}(0,1),
\\
\mathrm{Pr'}(0,0) &=  \mathrm{Pr}(0,0) .
\end{align}
\end{subequations}
We have introduced a prime notation to distinguish the case of number resolving detectors from non-number resolving detectors. The corresponding density matrices are
\begin{subequations}
\begin{align}
\rho'_{\qty(1,1)} &= \rho_{\qty(1,1)},
\\
\rho'_{\qty(1,0)} &= \frac{1}{\mathrm{Pr'}(1,0)}\qty[\mathrm{Pr}(2,0) \rho_{\qty(2,0)} + \mathrm{Pr}(1,0)\rho_{\qty(1,0)}],
\\
\rho'_{\qty(0,1)} &= \frac{1}{\mathrm{Pr'}(0,1)}\qty[\mathrm{Pr}(0,2) \rho_{\qty(0,2)} + \mathrm{Pr}(0,1)\rho_{\qty(0,1)}],
\\
\rho'_{\qty(0,0)} &= \rho_{\qty(0,0)}.
\end{align}
\end{subequations}
In this work we do not account for less-than-unity efficient detectors. However, this could easily be modelled by including an additional weighted BS in front of each detector and thus introduction of two new frequency-independent leakage channels.

\subsection{Identical monochromatic two-photon input}
\noindent
For identical photon envelopes and without any temporal displacement we have $ \xi(\omega,\omega')=\xi(\omega',\omega) $ and we find $ \delta\xi = 0 $. Furthermore, for monochromatic photons the bound state contribution from two-photon scattering vanishes. The physical argument is that the temporal spread in the photon, and hence the spread in energy, causes the emitter to stay in the ground state at all times. This means that the anharmonic emitter energy-level spacing cannot be exploited and only the linear terms give a non-zero contribution to the scattered envelope. This is true for any $ N $-photon Fock state. We drop the contributions from photon loss so the post scattering state in Eq.~\eqref{eq:2Ph_Psi3} simplifies to
\begin{align}
\begin{split}
\ket{\Psi_3} &= \frac{-1}{4} \qty(t^2_1(\omega) \ket{\uparrow} +\ket{\downarrow})\otimes \qty(\ket{\uparrow} + \ket{\downarrow}) \otimes   \qty(u^{\dagger})^2\ket{\varnothing}  +\frac{1}{4} \qty(\ket{\uparrow} + \ket{\downarrow})\otimes \qty(t^2_2(\omega)\ket{\uparrow} +  \ket{\downarrow}) \otimes   \qty(d^{\dagger})^2\ket{\varnothing} \label{eq:2ph_Psi3_noLoss_identical}
\end{split}.
\end{align}
After the second BS the state is
\begin{align}
\begin{split}
\ket{\Psi_4} &= \frac{1}{8} \qty[\qty(-t_1^2(\omega)+t_2^2(\omega))\ket{\uparrow, \uparrow} +\qty(-t_1^2(\omega)+1)\ket{\uparrow, \downarrow} + \qty(-1+t_2^2(\omega))\ket{\downarrow, \uparrow} +0\ket{\downarrow, \downarrow}] \otimes   \qty(u^{\dagger})^2\ket{\varnothing}
\\
&\quad  +\frac{1}{8} \qty[\qty(-t_1^2(\omega)+t_2^2(\omega))\ket{\uparrow, \uparrow} +\qty(-t_1^2(\omega)+1)\ket{\uparrow, \downarrow} + \qty(-1+t_2^2(\omega))\ket{\downarrow, \uparrow} +0\ket{\downarrow, \downarrow}] \otimes   \qty(d^{\dagger})^2\ket{\varnothing}
\\
& \quad +\frac{1}{4} \qty[\qty(t_1^2(\omega)+t_2^2(\omega))\ket{\uparrow, \uparrow} +\qty(t_1^2(\omega)+1)\ket{\uparrow, \downarrow} + \qty(1+t_2^2(\omega))\ket{\downarrow, \uparrow} +2\ket{\downarrow, \downarrow}]\otimes   u^{\dagger}d^{\dagger}\ket{\varnothing},
\end{split}
\end{align}
and from this result we can show that the probability of detecting the two photons with one detector is
\begin{align}
\mathrm{Pr}(2,0) = \mathrm{Pr}(0,2) = \frac{1}{32}\qty(\qty|-t_1^2(\omega)+t_2^2(\omega)|^2 + \qty|-t_1^2(\omega)+1|^2+\qty|-1+t_2^2(\omega)|^2),
\end{align}
the probability of a coincidence detection is
\begin{align}
\mathrm{Pr}(1,1) = \frac{1}{16}\qty(\qty|t_1^2(\omega)+t_2^2(\omega)|^2 + \qty|t_1^2(\omega)+1|^2+\qty|1+t_2^2(\omega)|^2+4). 
\end{align}
The respective concurrences are
\begin{align}
\mathcal{C}(2,0) = \mathcal{C}(0,2) &= \frac{2\qty|\qty(-t_1^2(\omega)+1)\qty(-1+t_2^2(\omega))|}{\qty|-t_1^2(\omega)+t_2^2(\omega)|^2 + \qty|-t_1^2(\omega)+1|^2+\qty|-1+t_2^2(\omega)|^2},
\\
\mathcal{C}(1,1) &= \frac{\qty|4\qty(t_1^2(\omega)+t_2^2(\omega))-2\qty(t_1^2(\omega)+1)\qty(1+t_2^2(\omega))|}{\qty|t_1^2(\omega)+t_2^2(\omega)|^2 + \qty|t_1^2(\omega)+1|^2+\qty|1+t_2^2(\omega)|^2+4}.
\end{align}
These results are strikingly similar in form to the single-photon protocol. The main difference here is that the transmission coefficients are raised to the second power. The optimal result is thus no longer found when $ t_1(\omega)=t_2(\omega)=-1 $ but instead when $ t^2_1(\omega)=t^2_2(\omega)=-1 $.

\section{\label{sec:nPhotons}$ N $-photon concurrence calculations}
In this section we generalize the result for a $ N=n+m $ identical monochromatic photon state input.
\subsection{\label{ssec:nPhotons_transport}$ N $-photon transport}
\noindent
The initial two-qubit state is again taken to be the separable pure state
\begin{align}
\ket{\phi_\text{in}} &= \frac{1}{2}\qty(\ket{\uparrow} +\ket{\downarrow})\otimes \qty(\ket{\uparrow} + \ket{\downarrow}).
\end{align}
The optical input state is now
\begin{align}
\ket{\psi_{\mathrm{in}}} = \frac{1}{\sqrt{n!m!}}\qty(u_1^{\dagger})^n\qty(d_1^{\dagger})^m\ket{\varnothing},
\end{align}
and after the first BS this becomes 
\begin{align}
\ket{\psi_2} = \sum_{k=0}^{N}f_{N,n;k}\qty(u_2^{\dagger})^k\qty(d_2^{\dagger})^{N-k}\ket{\varnothing}, \qquad N \equiv n+m,
\end{align}
with
\begin{align}
f_{N,n;k} = \frac{1}{\sqrt{n!(N-n)!}}\frac{1}{\sqrt{2^N}} \sum_{k_1=\mathrm{max}(0,k+n-N)}^{\mathrm{min}(n,k)}\qty(-1)^{k-k_1}\mqty(n \\ k_1)\mqty(N-n \\ k-k_1).
\end{align}
Assuming identical photons the post-scattering state becomes
\begin{align}
\begin{split}
\ket{\Psi_3} &= \sum_{k=0}^{N} \frac{1}{2} \qty(t_1^k \ket{\uparrow} +\ket{\downarrow})\otimes \qty(t_2^{N-k}\ket{\uparrow} + \ket{\downarrow}) \otimes f_{N,n;k}\qty(u^{\dagger})^k\qty(d^{\dagger})^{N-k}\ket{\varnothing} ,
\end{split}
\end{align}
and this evolves through the second BS to 
\begin{align}
\begin{split}
\ket{\Psi_4} &= \sum_{k=0}^{N} \frac{1}{2} \qty(t_1^k \ket{\uparrow} +\ket{\downarrow})\otimes \qty(t_2^{N-k}\ket{\uparrow} + \ket{\downarrow}) \otimes f_{N,n;k} \sum_{p=0}^{N} g_{N,k;p}\qty(u^{\dagger})^p\qty(d^{\dagger})^{N-p}\ket{\varnothing} ,
\end{split}
\end{align}
with
\begin{align}
g_{N,k;p} = \frac{1}{\sqrt{2^N}} \sum_{p_1=\mathrm{max}(0,p+k-N)}^{\mathrm{min}(k,p)}\qty(-1)^{k-p_1}\mqty(k \\ p_1)\mqty(N-k \\ p-p_1).
\end{align}
This state can be re-cast as
\begin{align}
\begin{split}
\ket{\Psi_4} &= \sum_{p=0}^{N}\qty[c_{N,n;p}^{\uparrow \uparrow} \ket{\uparrow ,\uparrow} + c_{N,n;p}^{\uparrow \downarrow} \ket{\uparrow, \downarrow} + c_{N,n;p}^{\downarrow \uparrow} \ket{\downarrow, \uparrow} + c_{N,n;p}^{\downarrow \downarrow} \ket{\downarrow, \downarrow}]\otimes \qty(u^{\dagger})^p\qty(d^{\dagger})^{N-p}\ket{\varnothing} ,
\end{split}
\end{align}
with
\begin{gather}
c_{N,n;p}^{\uparrow \uparrow} = \frac{1}{2} \sum_{k=0}^{N}f_{N,n;k} g_{N,k;p} t_1^k t_2^{N-k}, \quad c_{N,n;p}^{\uparrow \downarrow} = \frac{1}{2} \sum_{k=0}^{N}f_{N,n;k} g_{N,k;p} t_1^k ,
\\
c_{N,n;p}^{\downarrow \uparrow } = \frac{1}{2} \sum_{k=0}^{N}f_{N,n;k} g_{N,k;p} t_2^{N-k}, \quad  c_{N,n;p}^{\downarrow \downarrow } = \frac{1}{2} \sum_{k=0}^{N}f_{N,n;k} g_{N,k;p}.
\end{gather}


\subsection{\label{ssec:nPhotons_conc}$ N $-photon measurement}
\noindent
The probability of detecting $ p $ photons at $D_1$ and $ q = N-p $ photons at $D_2$ is
\begin{align}
\mathrm{Pr}\qty(p,q) = p!q!\qty[|c_{N,n;p}^{\uparrow \uparrow}|^2 + |c_{N,n;p}^{\uparrow \downarrow}|^2 + |c_{N,n;p}^{\downarrow \uparrow}|^2 + |c_{N,n;p}^{\downarrow \downarrow}|^2].
\end{align}
This heralds the two-qubit state
\begin{align}
\ket{\phi_{\qty(p,q)}}=\frac{c_{N,n;p}^{\uparrow \uparrow} \ket{\uparrow ,\uparrow} + c_{N,n;p}^{\uparrow \downarrow} \ket{\uparrow ,\downarrow} + c_{N,n;p}^{\downarrow \uparrow} \ket{\downarrow, \uparrow} + c_{N,n;p}^{\downarrow \downarrow} \ket{\downarrow ,\downarrow}}{\sqrt{|c_{N,n;p}^{\uparrow \uparrow}|^2 + |c_{N,n;p}^{\uparrow \downarrow}|^2 + |c_{N,n;p}^{\downarrow \uparrow}|^2 + |c_{N,n;p}^{\downarrow \downarrow}|^2}},
\end{align}
which, assuming photon-number resolving detectors, leads to the concurrence
\begin{align}
\mathcal{C}(p,q) = \frac{|-2c_{N,n;p}^{\downarrow \downarrow}c_{N,n;p}^{\uparrow \uparrow} +2c_{N,n;p}^{\uparrow \downarrow}c_{N,n;p}^{\downarrow \uparrow}| }{|c_{N,n;p}^{\uparrow \uparrow}|^2 + |c_{N,n;p}^{\uparrow \downarrow}|^2 + |c_{N,n;p}^{\downarrow \uparrow}|^2 + |c_{N,n;p}^{\downarrow \downarrow}|^2}.
\end{align}

\subsection{\label{ssec:nPhotons_plot}$ N $-photon plots}
\begin{figure}[t!]
	\begin{center}
		\includegraphics[scale=1]{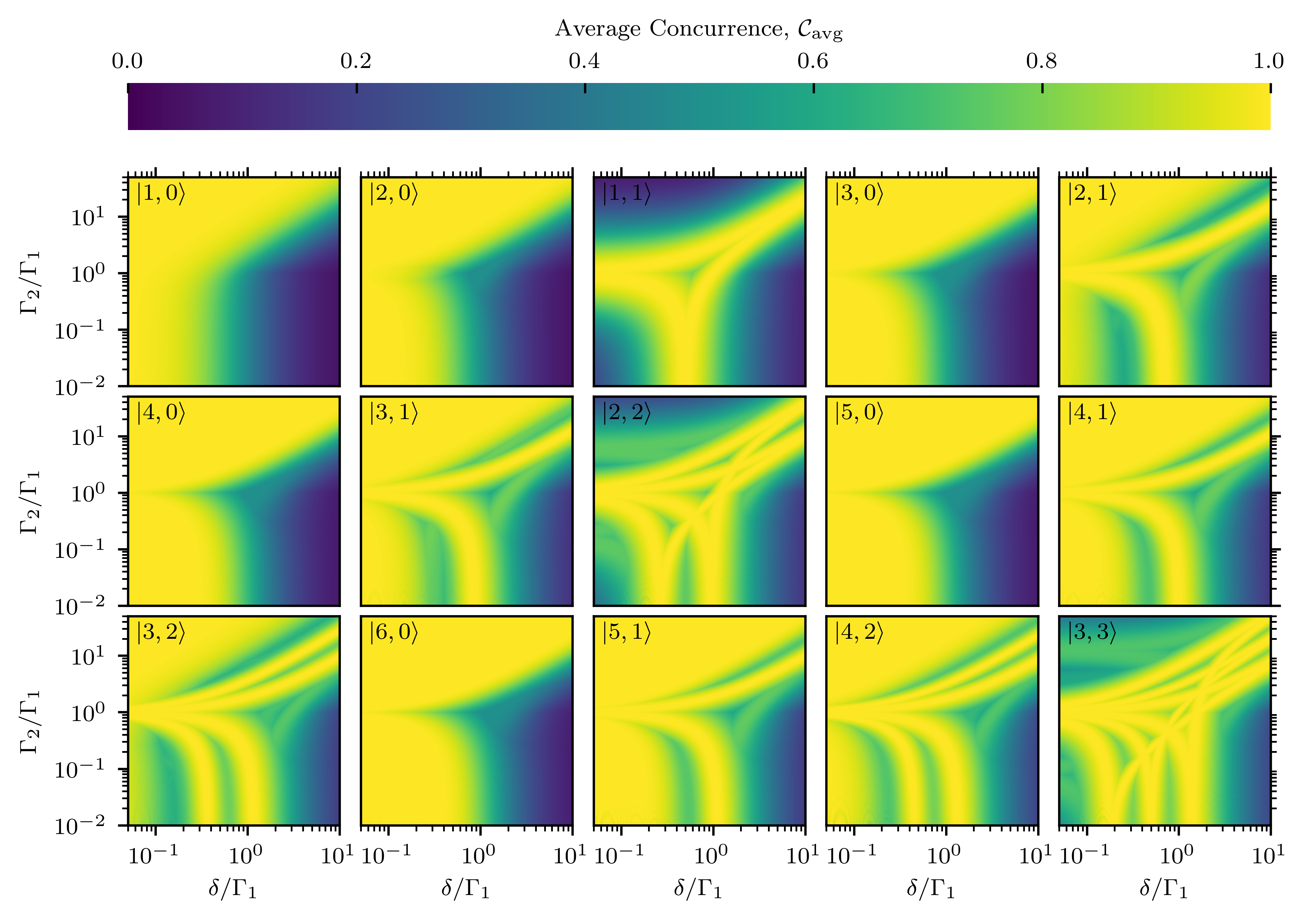}
		\caption{Plots of the average concurrence as a function of the coupling $\Gamma_2$ and emitter detuning $\delta$ for various input photon states $\ket{n,m}$.}
		\label{fig:figS1}
	\end{center}
\end{figure}

\noindent
In Fig.~\ref{fig:figS1} we plot the average concurrence as a function of the emitter properties, $ \Gamma_1 $, $ \Gamma_2 $, and $ \delta $, for up to $ 6 $ photons. We have assumed identical monochromatic photons and loss-less emitters.

\section{Optical State Optimisation}
\begin{figure}[t]
	\subfloat[Optimal photon frequency -- Single-photon]{\includegraphics[scale=0.9]{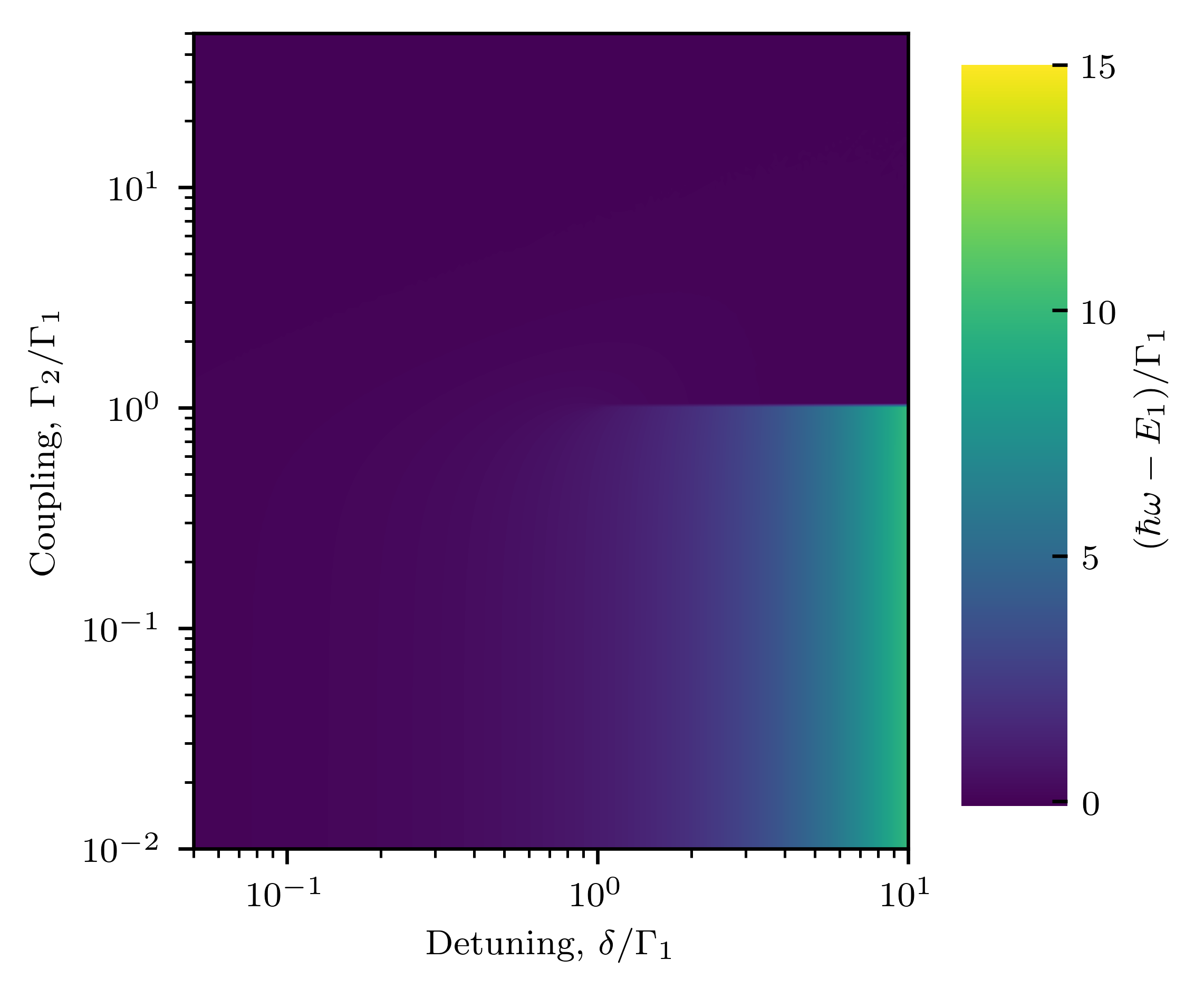}\label{fig:spopt}}
	\subfloat[Optimal photon frequency -- Two-photon]{\includegraphics[scale=0.9]{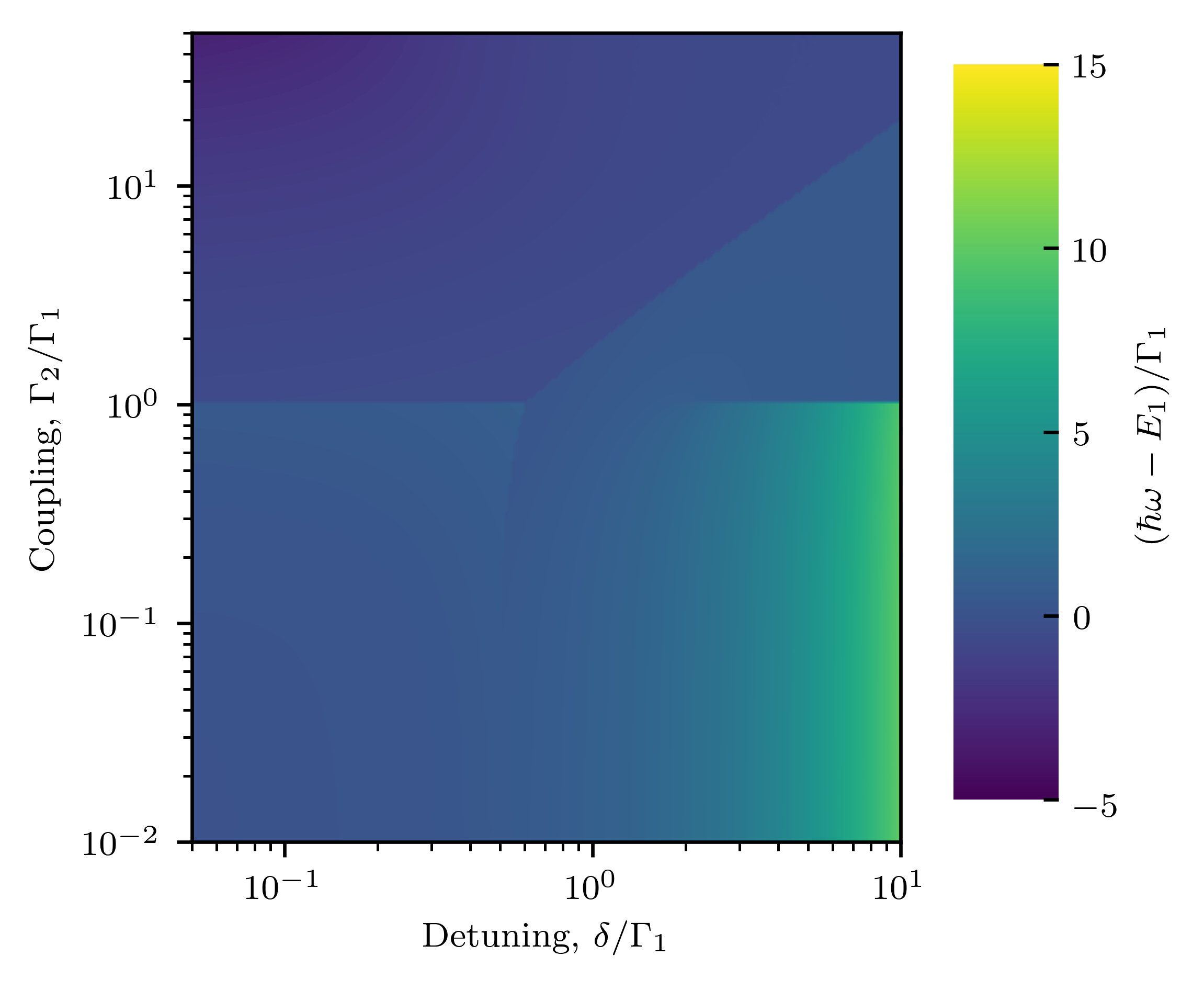}\label{fig:tpopt}}
	\caption{Energies of the photons in the input optical states that generate maximum possible entanglement between emitters coupled to the interferometer at rates $\Gamma_1$ and $\Gamma_2$ and detuned by $\delta$. In both the single, (a) and two, (b) photon case we assume a monochromatic optical source and lossless emitters. In the two-photon case we also assume that the photons are identical. }
	\label{fig:opteners}
\end{figure}

\noindent
For a given pair of emitters characterised by the coupling strengths $\Gamma_1$ and $\Gamma_2$ and spectrally separated by an energy $\delta$ we want to find the optical state that maximises the average concurrence generated between them. In Fig.~\ref{fig:opteners} we show the optimal energy of photons comprising one, \ref{fig:spopt} and two, \ref{fig:tpopt} photon optical input states. In both cases we assume that the states are monochromatic and the emitters are lossless. In the two-photon case we also assume that the photons are identical. We note several interesting features of the plots and first identify that at high detuning, $\delta$ and low coupling-strength the energy of the optimal input shifts with the detuning. This is because, as discussed in the main text, in this regime it becomes preferential to interact resonantly with one of the two emitters. We next see that for any other regime in the single photon case it is optimal to tune resonantly with the first emitter. This is because the spectrum of the second emitter overlaps well with the first, due either to its large width or the small separation between the spectra. In the two-photon case there is more subtle behaviour and at small spectral separations and high coupling-strengths we see that the optimal photon energy moves away from the first emitter. As discussed in the main text, in the monochromatic case the phase shift imparted on the emitters is additive as the number of photons increases. This means that we have to tune the optical state away from the emitters so that each photon imparts a phase of $\sim\pi/2$ radians so that the total shift is the required $\pi$ radians.

\bibliography{refs}